\documentclass[twocolumn,pre]{revtex4}
\usepackage{amsfonts}
\usepackage{amsmath}
\usepackage{amssymb}
\usepackage{graphicx}

\begin{document}

\title{Electrostatic effects in DNA stretching}
\author{Alexei V. Tkachenko (alexei@umich.edu)}
\affiliation{Department of Physics and Michigan Center for Theoretical Physics \\
University of Michigan, 450 Church Str., Ann Arbor, 48109 MI, USA}

\begin{abstract}
The response of a semiflexible polyelectrolyte chain to stretching in the
regimes of moderate and weak screening is studied theoretically, with a
special focus on DNA experiments. By using the nonlinear Poisson--Boltzmann
description of electrostatic self--interactions of the chain, we explicitly
demonstrate the applicability of the concept of effective charge to certain
aspects of the problem. This charge can be extracted from the far--field
asymptotic behavior of the electrostatic potential of the fully aligned
chain. Surprisingly, in terms of the effective charge, the electrostatically
renormalized persistence length can be formally described by the classical
Odijk-Skolnick-Fixman (OSF) formula, whose domain of applicability is
normally limited to linearized Debye Huckel (DH) approximation. However, the
short scale behavior of the chain in the nonlinear regime deviates from the
of DH--based result, even upon the charge renormalization. This difference
is revealed in the calculated stretching curves for strongly charged DNA.
These results are in good agreement with the recent experiments. In the
limit of weak screening we predict the elastic response to have a
distinctive two-stage character, with a peculiar intermediate
"unstretchable" regime.

\textbf{PACS:} 82.35.Rs, 87.14.Gg
\end{abstract}

\maketitle

\section{Introduction\protect\bigskip}

Single molecular micromechanical experiments are among the major tools of
modern biophysics. Pioneered by the studies of double--stranded DNA (dsDNA)%
\cite{DNAstretch}, these techniques are now applied to a wide range of
biologically relevant problems \cite{DNAstretch1}-\cite{DNAstretch3}.
Theoretically, it has been demonstrated that the observed response of dsDNA
to stretching (i.e. elongation vs. pulling force) can be successfully
described by semiflexible Worm-Like Chain (WLC) model \cite{kratky}\cite%
{Marko-Siggia}, over a considerable range of applied forces. Within this
model, the molecule is viewed as a rigid rod subjected to thermal
fluctuations, and the only free parameter is its persistence length $l_{p}$
which is proportional to the bending modulus of the rod.

Since DNA is strongly charged, the effects of the electrostatic interactions
should in principle play a significant role in its elastic response.
However, they are typically suppressed due to the strong screening under
physiological conditions. In addition, it is implicitly assumed that the
electrostatic effects can be incorporated by renormalization of the
parameters of the effective models, e.g. persistence length $l_{p}$ in WLC.
This is indeed a valid assumption in certain regimes, as was shown in works
of Odijk, and Skolnick and Fixman (OSF) \cite{OSF}. They have demonstrated
that within linearized Debye-Huckel (DH) approximation, there is an additive
electrostatic correction to the persistence length of a semi-flexible WLC,
and this term scales quadratically with Debye screening length $r_{s}$. \
However, the classical result has to be significantly revised in the regime
of highly flexible chains, which was in focus of numerous studies over the
last two decades \cite{KK}-\cite{polyel3}.

Another limitation of OSF picture is that it is only valid on the length
scales larger than $r_{s}$. When the chain is strongly stretched, its
properties on shorter scales are being probed. Within DH\ theory, this
regime can be described in terms \textit{scale-dependent} electrostatic
rigidity \cite{Barrat} \cite{Marko-Siggia}. Finally, OSF result is known to
be inadequate outside of DH approximation, i.e. in strongly nonlinear
regime. Compared to other aspects of the problem, its non-linear
generalization has attracted only a limited attention since the early works
back in '80s \cite{nonlin1} \cite{nonlin2}. In this paper, we study the
electrostatic effects on the elastic response of strongly charged WLC, in
the regimes of moderate and weak screening. In these regimes, DH
approximation may no longer be applicable, and the analysis of the
non-linear Poisson-Boltzmann equation is necessary. While stated as a
generic problem, our study is especially important in the context of DNA
stretching experiments.

The plan of the paper is as follows. In Section II, we formulate the problem
and review the results obtained within DH theory. We revisit the problem of
scale-dependent rigidity, and demonstrate that in the limit of weak
screening, the chain exhibits a two-stage response. These two stages are
characterized by the renormalized and bare persistence lengths,
respectively, and they are separated by a peculiar "unstretchable" regime.
In Section III, we study the problem beyond the linearized DH approximation.
Our central result is that the effective rigidity in the non-linear regime
can be described by classical OSF formula, in terms of \ \textit{effective
linear charge density} determined by the far-field asymptotic of chain's
electrostatic potential. This is the first explicit demonstration of the
applicability \ of the concept of renormalized charge in the context of \
the non-linear problem. While the notion of the effective charge is not new,
in the past \ it was introduced in a purely heuristic manner. As follows
from our results, the effective charge approach has significant limitations
and it is not valid on the length scale comparable to $r_{s}$. As a result,
the extension curve for strongly charged chains is different from the result
of linearized DH theory, even upon the charge renormalization. Finally, in
Section IV we discuss applications of the theory both to ds- and single
stranded (ss-)DNA, and compare our results to the existing experiments.

\section{Results of linearized theory}

In the presence of monovalent ions of concentration $n$, the electrostatic
potential in water is well described by Poisson-Boltzmann (PB) equation: 
\begin{equation}
\Delta \Psi \left( \mathbf{x}\right) -r_{s}^{-2}\sinh \Psi \left( \mathbf{x}%
\right) =-4\pi l_{B}\frac{\rho \left( \mathbf{x}\right) }{e}  \label{PB}
\end{equation}%
Here, reduced potential $\Psi $ is expressed in units of $kTD/e$; $D$ is the
dielectric constant of water; $l_{B}=e^{2}/DkT\approx 7\mathring{A}$ is
Bjerrum length , and $r_{s}=1/\sqrt{4\pi l_{B}n}$ is Debye screening length.
For low value of $\Psi $, the above non-linear equation can be linearized,
which corresponds to the well known DH approximation. In particular, one can
obtain Yukawa-type pair potential acting between two point charges $e$: 
\begin{equation}
\frac{V(r)}{kT}=\frac{l_{B}}{r}\exp \left( -r/r_{s}\right) .
\end{equation}

The effective Hamiltonian of an electrostatically neutral WLC stretched by
an external mechanical force is given by \cite{Marko-Siggia}: 
\begin{equation}
H_{0}=\int\limits_{0}^{L}\left[ \frac{kTl_{p}}{2}\left( \frac{\partial ^{2}%
\mathbf{x}}{\partial s^{2}}\right) ^{2}-\mathbf{f}\frac{\partial \mathbf{x}}{%
\partial s}\right] \mathrm{d}s
\end{equation}%
Here $\mathbf{f}$ is the stretching force applied at $s=L$. Function $%
\mathbf{x}\left( s\right) $ determines the chain conformation in space, and
it is subjected to constrain $\left\vert \partial \mathbf{x}/\partial
s\right\vert =1$ (i.e. the chain is not extendable). Within DH
approximation, the electrostatic self-interactions of uniformly charged
chain give rise to the following new term in Hamiltonian%
\begin{equation}
H_{el}=\frac{e^{2}\alpha ^{2}}{2Dl_{B}^{2}}\int\limits_{0}^{L}\int%
\limits_{0}^{L}\frac{\exp \left( -\left\vert \mathbf{x}\left( s\right) 
\mathbf{-x}\left( s^{\prime }\right) \right\vert /r_{s}\right) }{\left\vert 
\mathbf{x}\left( s\right) \mathbf{-x}\left( s^{\prime }\right) \right\vert }%
\mathrm{d}s\mathrm{d}s^{\prime }
\end{equation}%
Here $\alpha e/l_{B}$ is the linear charge density of the chain. Manning
parameter $\alpha $ has the physical meaning of the number of electron
charges per Bjerrum length.

According to OSF theory \cite{OSF} developed within DH approximation, the
electrostatic interactions lead to a higher effective bending modulus of a
semirigid chain. This results in renormalization of its persistence length: 
\begin{equation}
l_{p}^{\ast }=l_{p}+\alpha ^{2}r_{s}^{2}/4l_{B}.  \label{OSF}
\end{equation}%
For the case of flexible chain, OSF approach has been revised by Khokhlov
and Khachaturian, who have introduced the concept of electrostatic blobs 
\cite{KK}. On the length scales smaller than blob size, $\xi _{e}=\left(
l_{p}^{2}l_{B}/\alpha ^{2}\right) ^{1/3}$, the electrostatic effects may be
neglected.\ On the larger scales, a flexible polyelectrolyte can be viewed
as a stretched chain of blobs, and OSF picture can be recovered upon proper
renormalization of its parameters. For a semiflexible chains, we can
introduce a characteristic length which would play the same role as the size
of electrostatic blobs $\xi _{e}$ in flexible case: $l_{e}=DkT/(e\tau
)^{2}=l_{B}/\alpha ^{2}$. This length defines the minimal scale at which
electrostatic effects become relevant, and the semiflexible regime itself is
determined by condition $l_{e}\lesssim l_{p}$.

When the chain is strongly stretched, one may keep only terms quadratic in
the transverse components of displacement, $\mathbf{u}_{s}\mathbf{=x}\left(
s\right) \mathbf{-f}\left( \mathbf{f\cdot x}\left( s\right) \right) /\mathbf{%
f}^{2}$. After going to Fourier representation, $\mathbf{u}%
_{s}=L^{-1/2}\sum\limits_{q}\mathbf{u}_{q}\exp \left( \mathrm{i}qs\right) $,
the overall effective Hamiltonian $H\equiv $ $H_{0}+H_{el}$ can be written
as: 
\begin{equation}
\frac{H\left[ \mathbf{u}_{q}\right] }{kT}=const+\sum\limits_{q}\frac{q^{2}%
\mathbf{u}_{q}\mathbf{u}_{-q}}{2}\left[ l_{p}q^{2}+\Lambda _{el}\left(
q\right) +\frac{f}{kT}\right]  \label{free_en}
\end{equation}%
Here, all the electrostatic effects are collected within the term $\Lambda
_{el}\left( q\right) $, which has the physical meaning of \textit{%
scale-dependent electrostatic tension}: 
\begin{equation}
\Lambda _{el}\left( q\right) =\frac{\alpha ^{2}}{2l_{B}}\left[ \left( 1+%
\frac{1}{q^{2}r_{s}^{2}}\right) \log \left( 1+q^{2}r_{s}^{2}\right) -1\right]
\label{lambda}
\end{equation}%
This electrostatic term was originally introduced by Barrat and Joanny \cite%
{Barrat}, and later adopted to the case dsDNA in the classical paper of
Marko and Siggia \cite{Marko-Siggia}. By using the Equipartition Theorem,
they have calculated the difference between the end-to-end distance $R$ and
the length of the fully--stretched chain , $L$. 
\begin{equation}
\frac{L-R}{L}=-\frac{\left\langle \mathbf{\dot{u}}^{2}\right\rangle }{2}%
=\int\limits_{-\infty }^{+\infty }\frac{1}{l_{p}q^{2}+\Lambda _{el}\left(
q\right) +f/kT}\frac{\mathrm{d}q}{2\pi }  \label{result}
\end{equation}%
\begin{figure}[tbp]
{\includegraphics[
height=4.3in,
width=3.1in
]{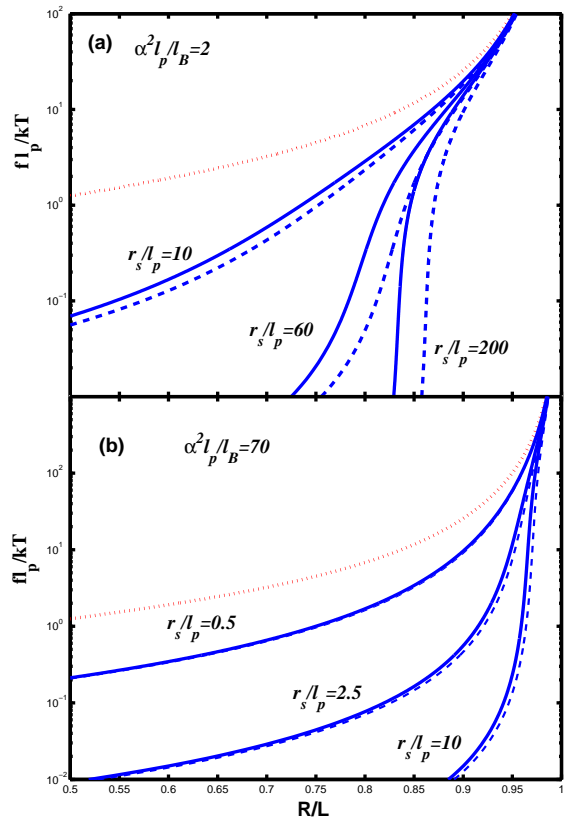}}
\caption{Stretching curves calculated for semiflexible polyelectrolytes at
various conditions. The solid lines are obtained within the linearized DH
approximation, while the dashed lines correspond to the complete nonlinear
PB description of the electrostatics. The stretching curve of a neutral
chain is shown with a dotted line. }
\label{force-length-fig}
\end{figure}

The calculated stretching curves are shown in Figure \ref{force-length-fig}.
Interesting observations can be made by further analysis of the above result
in the limit of weak enough screening, $r_{s}\gg l_{e}\equiv \alpha
^{-2}l_{B}$. In this case, the overall mechanical response has a striking 
\emph{two--stage} character. For large forces, $f\gg kT/4l_{e}$ we recover
well--known result for a neutral WLC \cite{Marko-Siggia}: 
\begin{equation}
\frac{L-R}{L}\simeq \int\limits_{-\infty }^{+\infty }\left[ l_{p}q^{2}+\frac{%
f}{kT}\right] ^{-1}\frac{\mathrm{d}q}{2\pi }=\frac{1}{2}\sqrt{\frac{kT}{%
fl_{p}}}  \label{flarge}
\end{equation}%
In addition, one can identify a new moderate force regime, $%
kTl_{e}/4r_{s}^{2}\lesssim f\ll kT/4l_{e}$. In this case, Eq. (\ref{result})
has two distinct contributions which correspond to the integration over wave
numbers $q$\ much smaller and much larger than $r_{s}^{-1}$, respectively:%
\begin{equation}
\frac{L-R}{L}\simeq \frac{1}{2}\sqrt{\frac{kT}{f\left( l_{p}+\alpha
^{2}r_{s}^{2}/4l_{B}\right) }}+\frac{1}{\alpha }\sqrt{\frac{l_{B}}{l_{p}}}%
g\left( \frac{\alpha r_{s}}{\sqrt{l_{B}l_{p}}}\right)  \label{fmoderate}
\end{equation}%
The first term here is similar to the interaction--free elastic response,
Eq. (\ref{flarge}), with the renormalized persistence length $l_{p}^{\ast
}=l_{p}+\alpha ^{2}r_{s}^{2}/4l_{B}$. This is consistent with OSF result,
Eq. (\ref{OSF}). The second term is independent of force, and it indicates
that stretching virtually stops when $kTl_{e}/4r_{s}^{2}\ll f\ll kT/4l_{e}$,
while $(L-R)/L$ remains finite due to fluctuations on the scales smaller
than $r_{s}$. We conclude that both the effective bending rigidity \emph{and
the total chain length} should be renormalized due to electrostatic
interactions. In other words, the chain becomes nearly "\emph{unstretchable"}
when end-to-end distance reaches certain value $R^{\ast }<L$ (as shown in
Figure \ref{force-length-fig}). Electrostatic effects are relevant only if $%
\alpha r_{s}\gtrsim \sqrt{l_{B}l_{p}}$. When this condition holds, function $%
g$ in Eq. (\ref{fmoderate}) has only weak (inverse-logarithmic) dependence
on the screening length: 
\begin{equation}
\frac{L-R^{\ast }}{L}\simeq \frac{1}{2\alpha \log \left( \alpha r_{s}/\sqrt{%
l_{B}l_{p}}\right) }\sqrt{\frac{l_{B}}{l_{p}}}  \label{R*}
\end{equation}%
The chain remains "unstretchable" until the force becomes strong enough to
suppress the small-scale fluctuations ( $f\sim kTl_{p}/r_{s}^{2}$). After
that, the crossover to large force regime (\ref{flarge}), occurs.

\section{\protect\bigskip Nonlinear theory and effective charge}

Strictly speaking, the linearization of Eq. (\ref{PB}) is only possible for
weekly charged chains or very strong screening. In a more general case, one
has to study the complete non-linear equation. In order to apply our results
for the case of strongly--charged polymers like ds- or ss- DNA, one has to
go beyond the linearized DH\ approximation. Early attempts of such a
generalization go back to early 80s \cite{nonlin1} \cite{nonlin2}. At that
time, it was shown by the numerical studies that effective rigidity clearly
deviates from the linear OSF theory. However, no alternative model had been
proposed.

Various aspects of PB\ equation in cylindrical geometry have been studied
for several decades. Among other important advances, was the discovery of
the counter--ions condensation by Manning \cite{Manning}. It was shown that
when the linear charge density on a thin rod exceeds certain critical value,
a finite fraction of the counter--ions gets localized within a near region
of the rod, thus effectively reducing its overall charge. This phenomenon
occurs at critical Manning parameter $\alpha _{c}\simeq 1$. Upon the
condensation, the residual effective charge of the rod saturates at its
critical value, $\alpha _{c}/l_{B}\simeq 1/l_{B}$. For a strongly--charged
cylinder with a finite radius $r_{0}$, a similar phenomenon occurs with the
effective charge becoming a function of aspect ratio, $r_{0}/r_{s}$.
Following Oosawa \cite{oosawa}, the idea of separating condensed and free
counter--ion components, has been widely used in literature. Nevertheless,
it is hard to justify its applicability for our problem. First, there is an
ambiguity in defining the "near" and "far" region of the cylinder with a
finite aspect ratio, and therefore the effective charge is not well defined.
Furthermore, the electrostatic interactions are strongest on lengths scales
shorter than $r_{s}$, exactly were DH approximation is invalid (for $\alpha
\gtrsim 1$).

The leading corrections to the free energy due to small fluctuation of the
chain can be evaluated without any model assumptions. Namely, one can
linearize PB equation near its solution for a static cylinder ($\Psi
^{\left( 0\right) }\left( \mathbf{x}\right) $), and couple the perturbations
of potential $\phi \left( \mathbf{x}\right) =\Psi \left( \mathbf{x}\right)
-\Psi ^{\left( 0\right) }\left( \mathbf{x}\right) $, to the conformational
fluctuations $\mathbf{u}_{s}$. \ The displacements of the charged cylinder
can be described by the local perturbations to the \textit{in-plane}
electrostatic moments. In particular, the monopole, dipole and quadrupole
moments have the following forms, respectively:%
\begin{equation}
\tau _{s}=\frac{\tilde{\alpha}e}{2l_{B}}\left( \frac{\partial \mathbf{u}_{s}%
}{\partial s}\right) ^{2};
\end{equation}%
\begin{equation}
\mathbf{d}_{s}=\frac{\tilde{\alpha}e}{l_{B}}\mathbf{u}_{s};
\end{equation}%
\begin{equation}
Q_{s}^{ij}=\frac{\tilde{\alpha}e}{l_{B}}\left[ \left( u_{s}^{i}u_{s}^{j}-%
\frac{\mathbf{u}_{s}^{2}\delta ^{ij}}{2}\right) +O\left( \left( r_{0}\frac{%
\partial \mathbf{u}_{s}}{\partial s}\right) ^{2}\right) \right] .
\end{equation}%
Here, $\tilde{\alpha}=\alpha _{0}+\left( r_{0}/2r_{s}\right) ^{2}\sinh \Psi
^{\left( 0\right) }\left( r_{0}\right) $ takes into account the fact that
the cylinder of radius $r_{0}$ with linear density density $e\alpha
_{0}/l_{B}$, fluctuates in oppositely--charged environment with local charge
density $ne\sinh \Psi ^{\left( 0\right) }\left( r_{0}\right) $. We have
limited our consideration to the leading three electrostatic moments because
they are the only ones containing displacement $\mathbf{u}_{s}$ in orders no
greater than $2$. Thus, the general expression for the correction to the
electrostatic Hamiltonian (second-order in $u$) is, 
\begin{equation}
\delta H_{el}=\int\limits_{0}^{L}\left[ \frac{\delta H_{el}}{\delta \tau _{s}%
}\tau _{s}-\frac{H_{el}^{\left( 0\right) }}{2L}\left( \frac{\partial \mathbf{%
u}_{s}}{\partial s}\right) ^{2}\right] \mathrm{d}s+  \label{H-el-nonlin}
\end{equation}%
\begin{equation*}
+\frac{1}{2}\int\limits_{0}^{L}\int\limits_{0}^{L}\left( \frac{\delta
^{2}H_{el}}{\delta \mathbf{d}_{s}\delta \mathbf{d}_{s^{\prime }}}\right) 
\mathbf{d}_{s}\mathbf{d}_{s^{\prime }}\mathrm{d}s\mathrm{d}s^{\prime }
\end{equation*}%
Here, the first term accounts for the fluctuations of the local linear
charge density, $\tau _{s}$. By symmetry reason, a similar term containing
quadrupole moment $\mathbf{\hat{Q}}_{s}$ is absent ($\delta H_{el}/\delta 
\mathbf{\hat{Q}}_{s}=0$). \ The second term in the above expression reflects
the fact that the cylinder is no longer perfectly straight and its
end-to-end distance is reduced by amount $\int\limits_{0}^{L}\left( \partial 
\mathbf{u}_{s}/\partial s\right) ^{2}ds/2$. The last term represents the
dipole-dipole interactions, and this is the only \textit{non-local} term in
the effective Hamiltonian:

\begin{equation}
\frac{1}{2}\int\limits_{0}^{L}\int\limits_{0}^{L}\left( \frac{\delta
^{2}H_{el}}{\delta \mathbf{d}_{s}\delta \mathbf{d}_{s^{\prime }}}\right) 
\mathbf{d}_{s^{\prime }}\mathbf{d}_{s^{\prime }}\mathrm{d}s\mathrm{d}%
s^{\prime }=
\end{equation}%
\begin{equation*}
=\frac{1}{D}\int\limits_{0}^{L}\int\limits_{0}^{L}\phi _{q}^{\left( 1\right)
}\left( r_{0},s-s^{\prime }\right) \left( \mathbf{d}_{s}-\mathbf{d}%
_{s^{\prime }}\right) ^{2}\mathrm{d}s\mathrm{d}s^{\prime }=
\end{equation*}%
\begin{equation*}
=\frac{2}{D}\sum\limits_{q}\left( \phi _{q}^{\left( 1\right) }\left(
r_{0}\right) -\phi _{0}^{\left( 1\right) }\left( r_{0}\right) \right) 
\mathbf{d}_{q}\mathbf{d}_{-q}
\end{equation*}%
Here we have switch to Fourier representation, and expressed the result in
terms of the dipolar field $\ \phi _{q}^{\left( 1\right) }\left( r\right) $.
The latter satisfies the inhomogeneous linearized PB equation: 
\begin{equation}
\left[ \partial _{r}^{2}+\frac{1}{r}\partial _{r}-\left( q^{2}+\frac{1}{r^{2}%
}+\frac{1}{r_{s}^{2}}\cosh \Psi ^{\left( 0\right) }\left( r\right) \right) %
\right] \phi _{q}^{\left( 1\right) }=0,  \label{PBdip}
\end{equation}%
subject to boundary condition:%
\begin{equation}
\left. \left( \frac{1}{r}-\partial _{r}\right) \phi _{q}^{\left( 1\right)
}\right\vert _{r=r_{0}}=1.
\end{equation}

Thus, 
\begin{equation}
\frac{\delta H_{el}\left[ \mathbf{u}_{q}\right] }{k_{B}T}=\sum\limits_{q}%
\left[ \frac{2\tilde{\alpha}^{2}}{l_{B}}\left( \phi _{q}^{\left( 1\right)
}\left( r_{0}\right) -\phi _{0}^{\left( 1\right) }\left( r_{0}\right)
\right) \mathbf{u}_{q}\mathbf{u}_{-q}+Cq^{2}\mathbf{u}_{q}\mathbf{u}_{-q}%
\right]  \label{delta H}
\end{equation}%
Here the we have used the fact that the first two terms in Eq. (\ref%
{H-el-nonlin})\ are local in $\partial \mathbf{u}_{s}/\partial s$: 
\begin{equation*}
const\cdot \int \left( \partial \mathbf{u}_{s}/\partial s\right)
^{2}ds=\sum\limits_{q}Cq^{2}\mathbf{u}_{q}\mathbf{u}_{-q}
\end{equation*}%
We can determine coefficient $C$ by noting that the low-$q$ behavior of the
electrostatic correction to the Hamiltonian, $H_{el}\left[ \mathbf{u}_{s}%
\right] $ should still be dominated by bending modes, $q^{4}\mathbf{u}_{q}%
\mathbf{u}_{-q}$, and therefore \ all the terms containing $q^{2}\mathbf{u}%
_{q}\mathbf{u}_{-q}$ must cancel in that limit:%
\begin{equation}
C=-\frac{\tilde{\alpha}^{2}}{l_{B}}\left( \left. \frac{\partial ^{2}\phi
_{q}^{\left( 1\right) }\left( r_{0}\right) }{\partial q^{2}}\right\vert
_{q=0}\right)
\end{equation}

After substitution of this value of $C$ into Eq. (\ref{delta H}), we obtain
a nonlinear generalization of the electrostatic tension term $\Lambda
_{el}\left( q\right) $ which enters \ the earlier results, Eqs. (\ref%
{free_en}) and (\ref{result}): 
\begin{equation}
\Lambda _{el}^{\left( nonlin\right) }\left( q\right) =\frac{4\tilde{\alpha}%
^{2}}{l_{B}}\left( \frac{\phi _{q}^{\left( 1\right) }\left( r_{0}\right)
-\phi _{0}^{\left( 1\right) }\left( r_{0}\right) }{q^{2}}-\left. \frac{%
\partial ^{2}\phi _{q}^{\left( 1\right) }\left( r_{0}\right) }{2\partial
q^{2}}\right\vert _{q=0}\right)   \label{LambdaPB}
\end{equation}%
\begin{figure}[tbp]
{\includegraphics[
height=2.2in,
width=3.1in
]{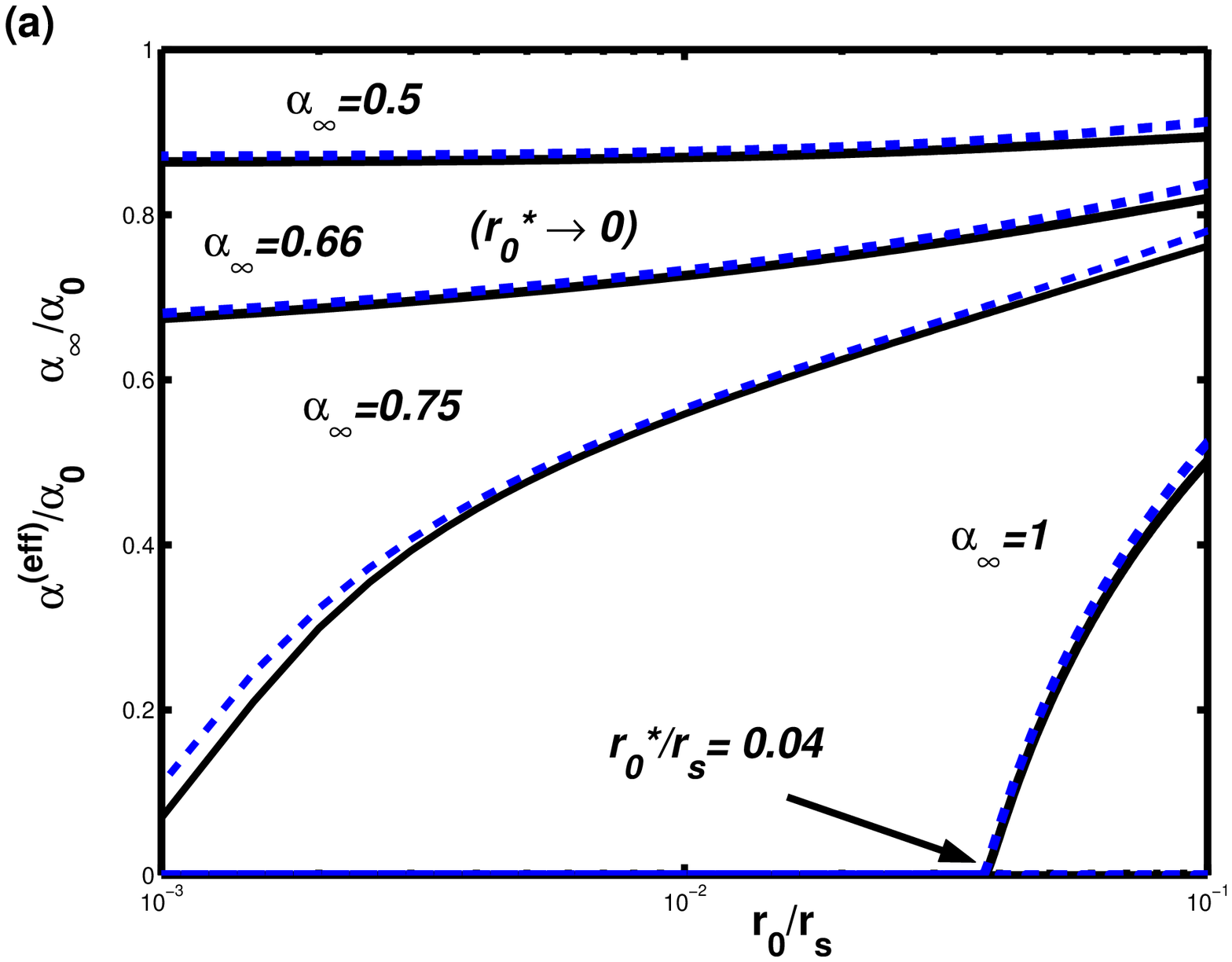}} {\includegraphics[
height=2.2in,
width=3.1in
]{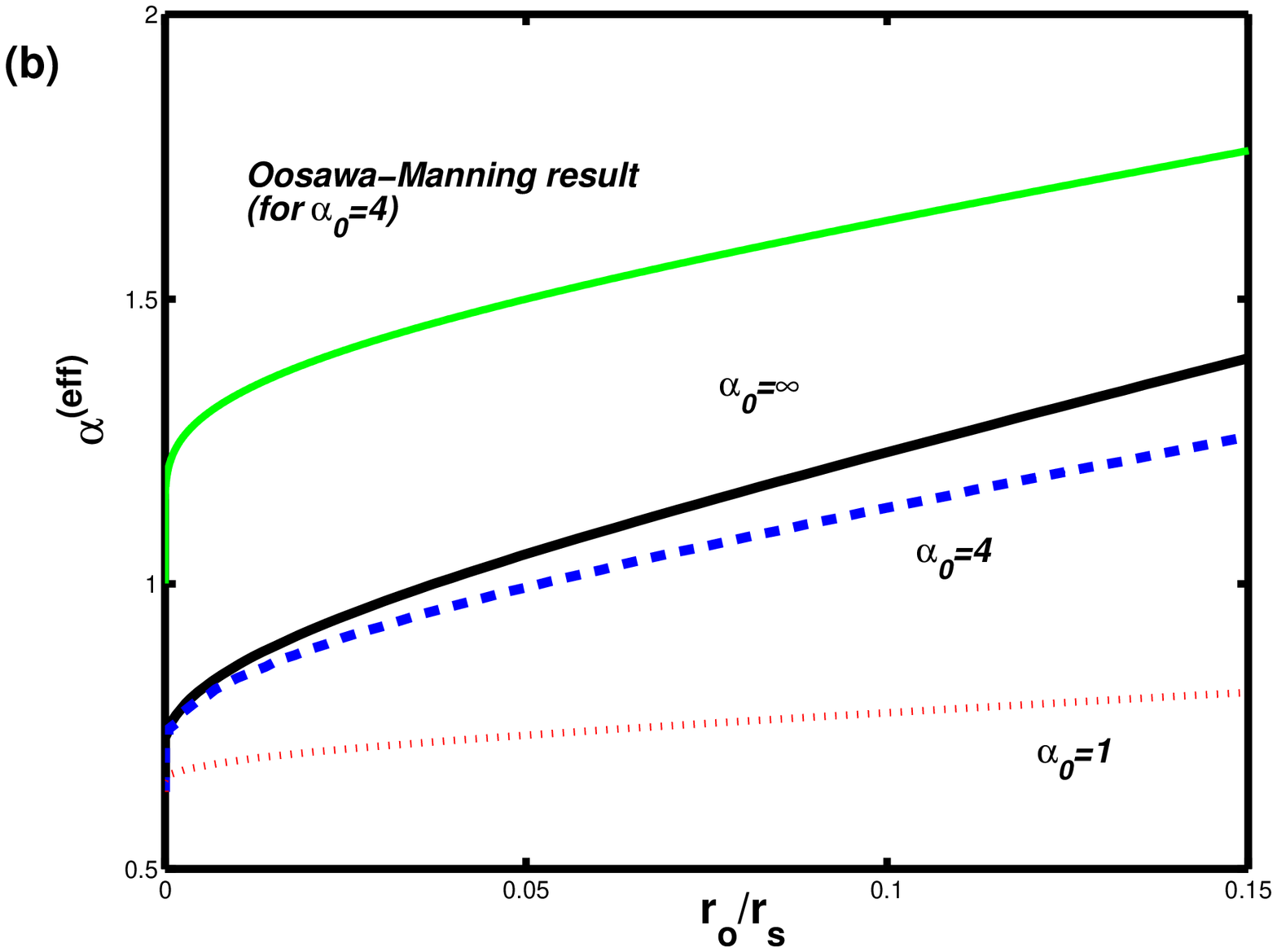}} {\includegraphics[
height=2.2in,
width=3.1in
]{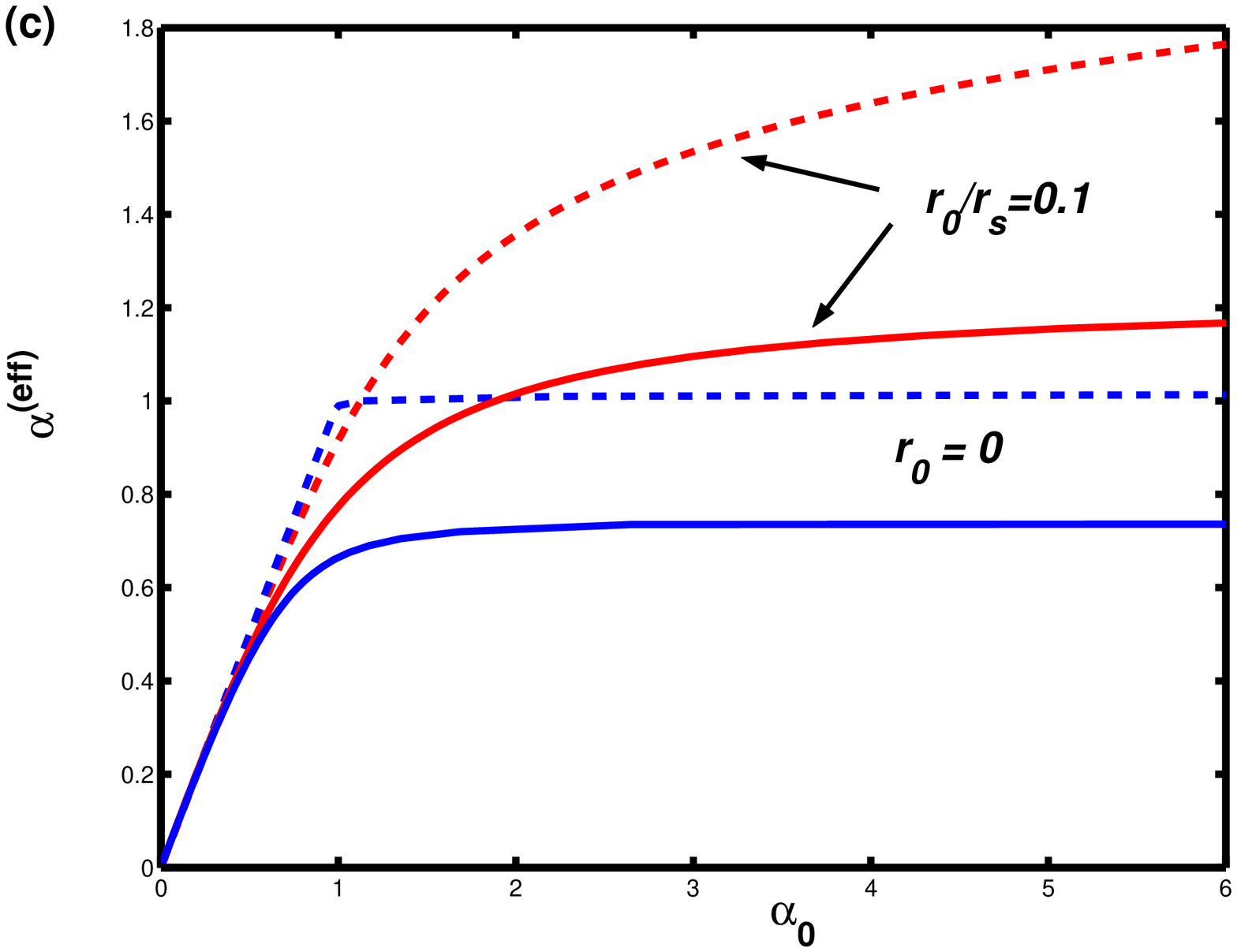}}
\caption{(a) Correlation between the effective Manning parameter $\protect%
\alpha ^{\left( eff\right) }$ obtained from the renormalized rigidity of the
strongly charged chain, and the one extracted from the far-field behavior of
the electrostatic potential, $\protect\alpha _{\infty }$. \ Solid and dashed
line represent $\protect\alpha ^{\left( eff\right) }/$ $\protect\alpha _{0}$
, and $\protect\alpha _{\infty }/$ $\protect\alpha _{0}$ , respectively, for
the fixed value of $\protect\alpha _{\infty }$ and variable aspect ratio $%
r_{0}/r_{s}$ . (b) $\protect\alpha ^{\left( eff\right) }$ as a function of
aspect ratio for various values of bare charge: $\protect\alpha %
_{0}\rightarrow \infty $ ( the saturation curve, solid line), $\protect%
\alpha _{0}=4$ (dsDNA, dashed line) and $\protect\alpha _{0}=1$ (ssDNA,
dotted line). Note that $\protect\alpha ^{\left( eff\right) }$ clearly
devites from the result obtained through Oosawa-type calculation, Ref. 
\protect\cite{joanny}. (c) Effective vs. bare charge dependence (solid
lines), compared to similar result of Oosawa-Manning model (dashed lines).
The two sets of plots correspond to the limit of an infinitely thin rod ($%
r_{0}=0$), and a finite aspect ratio ($r_{0}/r_{s}=0.1$), respectively. }
\label{Qeff-fig}
\end{figure}

By solving Eqs. (\ref{PB}) and (\ref{PBdip}) numerically, we found function $%
\Lambda _{el}^{\left( nonlin\right) }\left( q\right) $ for various regimes.
Its low-$q$ asymptotic may significantly deviate from classical OSF result,
which is consistent with the earlier studies \cite{nonlin1} \cite{nonlin2}.
One could interpret this deviation in terms of renormalized effective
charge, by defining the effective Manning parameter $\alpha ^{(eff)}$ so
that in the limit $qr_{s}\ll 1$,%
\begin{equation}
\Lambda _{el}^{\left( nonlin\right) }\left( q\right) \simeq \frac{\left(
\alpha ^{(eff)}r_{s}q\right) ^{2}}{4l_{B}}
\end{equation}%
Surprisingly, this effective effective charge appear to have a clear
physical meaning. With an excellent accuracy, $\alpha ^{(eff)}$ has\ the
same value as effective Manning parameter $\alpha _{\infty }$ which can be
extracted from the far-field behavior of potential $\Psi ^{\left( 0\right)
}\left( r\right) $, i. e. 
\begin{equation*}
\Psi ^{\left( 0\right) }\left( r\right) \simeq 2\alpha _{\infty
}K_{0}(r/r_{s})
\end{equation*}%
for $r\gtrsim r_{s}$ \cite{Qeff-farfield}. Here $K_{0}$ is the modified
Bessel function which is the solution to DH equation in cylindric geometry.

Figure \ref{Qeff-fig}(a) demonstrates a striking correlation between $\alpha
^{(eff)}$ and $\alpha _{\infty }$, in a wide range of parameters. Each curve
there corresponds to the fixed value of $\alpha _{\infty }$. The solid and
dashed lines represent $\alpha ^{(eff)}/\alpha _{0}$ and $\alpha _{\infty
}/\alpha _{0}$ respectively, as functions of the aspect ratio $r_{0}/r_{s}$.
The close agreement between the two types of data shows that $\alpha ^{(eff)}
$ and $\alpha _{\infty }$ nearly coincide for an arbitrary value of the bare
linear charge, as long as the aspect ratio $r_{0}/r_{s}$ remains small.
Figure \ref{Qeff-fig}(b) illustrates the dependence of $\alpha ^{(eff)}$ on
the aspect ratio for the values of the bare charge close to those of dsDNA
and ssDNA ($\alpha _{0}=4$ and $\alpha _{0}=1$, respectively). One can see
that for the practical purposes, $\alpha ^{(eff)}$ of ssDNA is close to $.7$%
, while its value for dsDNA follows the saturation curve (corresponding to
infinitely--charged cylinder with the same radius). Note that the far-field
definition of the effective charge is unambiguous, and it does deviate from
Oosawa--Manning result, even for zero aspect ratio (see Figure \ref{Qeff-fig}%
(c)) . Furthermore, one can modify Oosawa-Manning approach to evaluate the
effective charge as a function of $r_{0}/r_{s}$, as demonstrated in Ref. 
\cite{joanny}. The result of that calculation is also shown in Figures \ref%
{Qeff-fig}(b)-(c),  and it clearly deviates from our far-field based
definition of $\alpha ^{(eff)}$.

\begin{figure}[tbp]
{\includegraphics[
height=2.2in,
width=2.9in
]{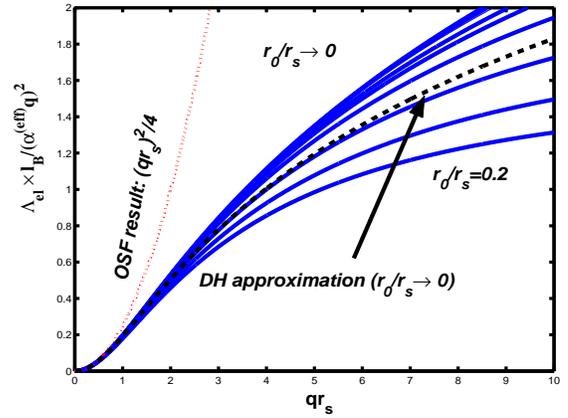}}
\caption{$\Lambda _{el}\left( q\right) $ obtained from the complete PB
equation, for various aspect ratios (solid lines). An introduction of the
effective Manning parameter results in matching of these curves with DH
(dashed), and OSF (dotted) results, for low $q$. For $qr_{s}\gtrsim 1$, the
DH approximation underestimates the effect of electrostatic interactions
(compare the zero aspect ratio curves). Both PB and DH curves significantly
deviate from OSF parabola, indicating that the effective rigidity
description is not valid on length scales smaller than $r_{s}$.}
\label{PB-fig}
\end{figure}
We have \emph{explicitly} demonstrated that OSF result is formally
applicable even outside of DH approximation, given that the effective charge
is properly defined. While similar effective charge descriptions have been
previously utilized in several works \cite{Qeff-farfield}\cite{zhang}, they
were all introduced as purely empirical assumption. In fact, our result is
rather non--trivial since the electrostatic interactions are strongest in
the vicinity of the chain, where the effective description is not valid. The
effective charge picture works only because the bending of the chain affects
the far region much stronger than the near one. Indeed, $q^{2}$ term in Eq. (%
\ref{PBdip}) is most relevant for $r\gtrsim r_{s}$.

Furthermore, for $q\gtrsim 1/r_{s}$ the effective charge description breaks
down, and deviations from DH approximation become considerable, as shown in
Figure \ref{PB-fig}. Qualitatively, this can be interpreted as an increase
of the effective charge with $q$: as one goes to smaller length scales, $%
\alpha ^{(eff)}$ should shift from its far-field value $\alpha _{\infty }$,
towards the bare Manning parameter $\alpha _{0}$. The change of the overall
shape of $\Lambda _{el}\left( q\right) $ gives rise to modification of the
stretching curves, compared to the results of DH--based theory, as shown in
Figure \ref{force-length-fig}. Naturally, these corrections are strongest
for large ratios $r_{s}/l_{p}$, which corresponds to wider dynamic range of
relevant wave vectors.

\section{Discussion}

The above results can be applied to the experiments on both ds- an ssDNA.
The two cases are characterized by rather different values of the physical
parameters. In particular, the persistence length of dsDNA, $l_{p}^{\left(
ds\right) }\simeq 50\mathrm{nm}\simeq 70l_{B}$ significantly exceeds that of
ssDNA chain, $l_{p}^{\left( ss\right) }\sim 1nm\sim l_{B}$. In addition,
their linear charge density differ approximately by factor of $4$. Finally,
while WLC model is well established as a standard description of dsDNA, the
situation with elasticity of ssDNA is not as settled: the extensible version
of WLC is among several proposed models all of which are in reasonable
agreement with the existing experimental data \cite{pnels}. Plots in Figures %
\ref{force-length-fig}(a) and (b) roughly correspond to the parameters of
ds- and ssDNA, respectively. As one can see \ from the Figure, the two-stage
behavior and the intermediate "unstretchable" regime becomes pronounced when 
$r_{s}$ substantially exceeds the persistence length. Note however that this
regime is unlikely to be accessible in the case of dsDNA, since the double
helix would become unstable at such a week screening due to the
electrostatic repulsion between the strands. For ssDNA, the two-stage
stretching must be observable at $r_{s}\gtrsim 100\mathrm{nm}$.

\begin{figure}[tbp]
{\includegraphics[
height=2.4in,
width=3.1in
]{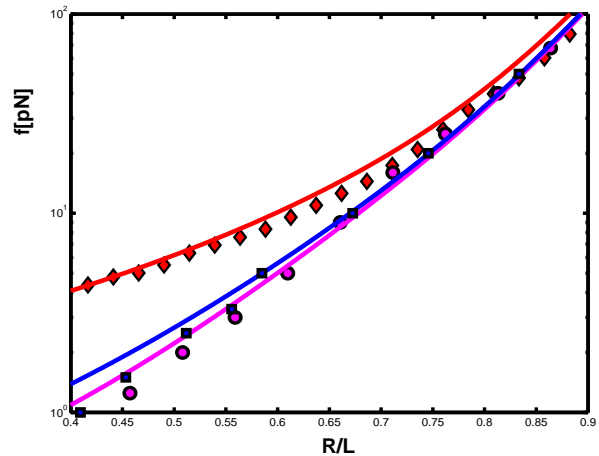}}
\caption{Comparison of the theoretical results to experimental data on
elastic response of ssDNA at various screening conditions. Data points are
taken from Refs. \protect\cite{DNAstretch1} (2mM NaCl, shown by circles),
and \protect\cite{DNAstretch3} (1mM and 10mM PB data are represented by
squares and diamonds, respectively). The theoretical curves were obtained by
taking into account the calculated values of Debye length for each case ( $%
6.9\mathrm{nm}$, $6\mathrm{nm}$ and $1.9\mathrm{nm}$ respectively). ssDNA
was approximated by a cylinder of radius $r_{0}=0.5\mathrm{nm}$. The only
fitting parameter for all three curves is bare persistence length $l_{p}$,
whose optimal value is close to $5.5\mathrm{nm}$. }
\label{exper}
\end{figure}

While Figure \ref{force-length-fig} gives only a general idea about the
mechanical response of semiflexible chain with electrostatic interactions, a
more detailed comparison of our results \ with experiments is presented in
Figure \ref{exper}. The plot shows the stretching \ curves of ssDNA taken at
various screening conditions, as reported in Refs. \cite{DNAstretch1} and 
\cite{DNAstretch3}. These experimental data are fitted by our theory with
the Debye screening length calculated for each salt concentration. As a
result, the only fitting parameter was \ the \textit{bare} persistence
length $l_{p}$, and the best fit is achieved at $l_{p}=5.5\mathring{A}$.
This value is somewhat smaller that the earlier estimates of the \textit{%
effective } persistence length $l_{p}^{\ast }\simeq 8\mathring{A}$, which
difference is due to our explicit account for the electrostatic corrections
to chain rigidity. Unlike the plots in Figure \ref{force-length-fig} which
for the sake of simplicity were calculated by assuming $r_{0}/r_{s}=0$, in
Figure \ref{exper} the non zero diameter of ssDNA chain is taken into
account. Based on the known structure of the molecule, we have assumed $%
r_{0}=5\mathring{A}$, and this parameter was not used for fitting.

Even though the theoretical stretching curves were calculated with WLC
model, our approach can be also applied to alternative descriptions of
ssDNA, such as \ Extensible Freely-Joint Chain model (EFJC). However, this
modification is unlikely to produce any significant change since within the
force range of interest, the competing models (WLC, EFJC, and discreet
version of WLC) are known to give nearly identical fitting to the
electrostatics--free stretching behavior of ssDNA \cite{pnels}. Indirectly,
this model independence is supported by the fact that the above experimental
data were also successfully fitted with numerical simulations based on EFJC
model \cite{DNAstretch3}. In that work, in addition to electrostatic
interactions described within linear DH approximation with renormalized
charges, the effects of excluded volume and hairpin formation were included 
\cite{zhang}. Since these additional effects are only important for relative
small chain extension, they were neglected in the context of our work.

\section{Conclusions}

In conclusion, we have studied the \ effects of electrostatic
self-interactions on elastic properties of strongly-stretched WLC, with
application to ds- and ss-DNA at moderate and weak screening conditions. By
revisiting the earlier studied problem of scale--dependent electrostatic
rigidity we have demonstrated that in the limit of weak screening, the
mechanical response of the chain has a distinctive two-stage character. The
two modes are characterized by the bare and the renormalized persistence
length, respectively, and they separated by intermediate "unstretchable"
regime. \ 

The central theme of our work is the effect of non-linearity of PB equation
on the classical results obtained within DH\ approximation. Consistently
with previous studies of the problem \cite{nonlin1} \cite{nonlin2}, the OSF
result for the electrostatic renormalization of the persistence length is
not applicable in the nonlinear regime, and in particular the scaling $%
l_{p}^{\ast }-l_{p}\sim r_{s}^{2}$ is not valid. However, we have explicitly
demonstrated that the OSF formula is formally correct if the actual line
charge density is replaced with the effective one. The latter is determined
from the far--field asymptotic of the electrostatic potential of the
strongly stretched chain. In the light of this result, the deviation from
OSF scaling is due to the non-trivial dependence of the effective charge on
the \ aspect ratio $r_{0}/r_{s}$.

The renormalized OSF description is however only partially adequate since
the electrostatic corrections to the chain rigidity do not follow the result
of DH theory on scales shorter that $r_{s}$, even upon the charge
renormalization. This deviation is due to the fact that the effective charge
itself is a scale-dependent concept. Our solution to the complete nonlinear
problem is shown to give an adequate description to the experimentally
observed stretching behavior of ssDNA at various salt conditions.

\begin{acknowledgments}
\textbf{Acknowledgments.} The author thanks S. Safran, Y. Rabin, H. Diamant
for valuable discussions, and US Department of State for ensuring a relaxed
and thought-provoking environment essential for this work.
\end{acknowledgments}

\end{document}